\documentclass[a4paper]{article}

\usepackage{graphicx}
\usepackage{amsmath}
\usepackage{array}
\usepackage{subfig}
\usepackage{layaureo}

\begin{document}

\title{An exponential correction to Starobinsky's inflationary model}
\author{J\'ulio C. Fabris\footnote{fabris@pq.cnpq.br}\;, Tays Miranda\footnote{taysmiranda91@gmail.com}\; and Oliver F. Piattella\footnote{oliver.piattella@cosmo-ufes.org}\\
Departamento de F\'{\i}sica - Universidade Federal do Esp\'{i}rito Santo, Brazil}

\titlepage
\maketitle

\centerline{\bf Abstract}

We analyse $f(R)$ theories of gravity from a dynamical system perspective, showing how the $R^2$ correction in Starobinsky's model plays a crucial role from the viewpoint of the inflationary paradigm. Then, we propose a modification of Starobinsky's model by adding an exponential term in the $f(R)$ Lagrangian. We show how this modification could allow to test the robustness of the model by means of the predictions on the scalar spectral index $n_s$.

\section{Introduction}

Inflation is a primordial phase of accelerated, quasi-exponential expansion of the universe required in order to solve the flatness and horizon problems, which plague standard Friedmannian cosmology. In most of the models, inflation is driven by a self-interacting scalar field $\phi$, known as {\it inflaton}, whose origin should be related to fundamental theories. See e.g. Ref.~\cite{mukha}.

The condition for which inflation is triggered is called \textit{slow roll} and it is parametrised by two \textit{slow-roll parameters} defined as follows:
\begin{eqnarray}\label{Slow}
\epsilon \equiv \frac{M_{Pl}^2}{2}\biggr(\frac{V'}{V}\biggl)^2\;, \quad  \eta \equiv  M_{Pl}^2\frac{V''}{V}\;,
\end{eqnarray}
where $M_{Pl}$ is the Planck mass, $V(\phi)$ is the potential term, and the prime denotes derivative with respect to $\phi$. When $\epsilon, \eta \ll 1$ then the condition $\dot\phi^2 \ll V$ is attained, i.e. the scalar field kinetic term is much smaller than the potential one. Therefore, the scalar field energy density is dominated by $V(\phi)$, which is almost constant, providing thus the accelerated, quasi-exponential expansion required.

Cosmological observables such as the spectral index of scalar and tensor perturbations are directly related to the slow-roll parameters:
\begin{eqnarray}\label{param}
n_s = 1 - 6\epsilon + 2\eta\;, \quad n_T = - 2\epsilon\;.
\end{eqnarray}

The recent cosmological observational data restrict considerably the viable inflationary models \cite{martin,planck1,planck2,tsujikawa}. One of the strongest candidate to describe the inflationary scenario is the Starobinsky model \cite{star1,star2,star3}. In its original form, it is based on a non-linear Lagrangian inspired from quantum effects and it can be viewed as an $f(R)$ model with $f = R + \alpha R^2$, i.e. a quadratic correction to the Einstein-Hilbert term.
 
Here we analyse the Starobinsky model from a dynamical system perspective and propose a generalisation that may lead to new interesting features. 


\section{Inflation from $f(R)$ theories and\\ Starobinsky's model}

We start showing why the $R^2$ correction of the Starobinsky model is somewhat special. Consider a generic $f(R)$ theory \cite{F}:
\begin{eqnarray}\label{frtheory}
	S = \frac{1}{2\kappa^2}\int d^4x\sqrt{-g}f(R)\;,
\end{eqnarray}
and assume a flat FLRW metric: 
\begin{eqnarray}
	ds^2 = -dt^2 + a(t)^2\delta_{ij}dx^idx^j\;.
\end{eqnarray}
One can easily cast the evolution equation for $f(R)$ gravity as a dynamical system formed by the following equations:
\begin{equation}\label{RdotHdot}
	\begin{cases}
		\dot{R} = \frac{1}{6Hf''}\left(Rf' - f - 6f'H^2\right)\;, \\
        \dot{H} = \frac{R}{6} - 2H^2\;,
	\end{cases}
\end{equation}
where $H \equiv \dot{a}/a$, the dot denoting derivation with respect to the cosmic time and the prime denoting derivation with respect to $R$. Here, we assumed $f''$ and $H$ different from zero.  

For a review of the techniques for the analysis of a dynamical system, see Ref. \cite{ds}. Looking for the critical points of the above autonomous system, i.e. setting to zero both the above equations and solving them simultaneously, one obtains the following ordinary differential equation for $f$:
\begin{eqnarray}
	\frac{f'}{f} = \frac{2}{R}\;,
\end{eqnarray}
which has the solution
\begin{eqnarray}
	f(R) = \alpha R^2\;,
\end{eqnarray}
where $\alpha$ is an integration constant. Here it appears the quadratic term $R^2$ as a special case of $f(R)$ theories for which the original dynamical system can be written as:
\begin{eqnarray}
	\begin{cases}
		\dot{R} = \frac{R}{12H}\left(R - 12H^2\right)\;,\\
\dot{H} = \frac{R}{6} - 2H^2\;.
	\end{cases}
\end{eqnarray}
We can now study the nature of the critical points. Linearising the dynamical system about a critical point $R_0 = 12H_0^2$, we obtain:
\begin{eqnarray}
	\begin{cases}
		\dot{\epsilon} = H_0\epsilon - 24H_0^2\eta\;,\\
\dot{\eta} = \frac{\epsilon}{6} - 4H_0\eta\;.
	\end{cases}
\end{eqnarray}
The secular equation for the above system matrix is $\lambda(\lambda + 3H_0) = 0$. Therefore, one eigenvalue is vanishing and the other is negative (for $H_0 > 0$, which is the region of our interest). This means that the critical points are attractors. The phase diagram is displayed in figure \ref{R2}. Physically, in this model Inflation can happen at any energy and this of course is not satisfactory because we want it to take place only at primordial times. 

Here it comes Starobinsky's model \cite{star1, star2, star3}, defined by $f(R) = R + \alpha R^2$. The linear, Einstein-Hilbert term $R$ breaks the parabola of critical points of the $R^2$ theory, leaving just a repulsive one at infinity, which represents the inflationary phase, and a critical point at finite distance. We show the latter explicitly. The dynamical system for this case becomes:
\begin{eqnarray}
	\begin{cases}
		\dot{R} = \frac{1}{12H\alpha}\left[\alpha R^2 - 6H^2(1 + 2\alpha R)\right]\;,\\
	\dot{H} = \frac{R}{6} - 2H^2\;,
	\end{cases}
\end{eqnarray}
in which $\alpha$ can be arranged as follows:
\begin{eqnarray}
	\begin{cases}
		\alpha^{3/2}\dot{R} = \frac{1}{12H\sqrt{\alpha}}\left[\alpha^2 R^2 - 6\alpha H^2(1 + 2\alpha R)\right]\;,\\
	\alpha\dot{H} = \frac{\alpha R}{6} - 2\alpha H^2\;,
	\end{cases}
\end{eqnarray}
in order for the dimensionless quantities $\alpha R$, $\alpha H^2$ and $t/\sqrt{\alpha}$ to appear.

Looking for critical points at finite distance from the origin in the phase-space $(H,R)$, we put $\dot{H} = 0$, obtaining thus $R = 12H^2$ implying:
\begin{eqnarray}
	\begin{cases}
		\dot{R} = -\frac{H}{2\alpha}\;,\\
	\dot{H} = 0\;.
	\end{cases}
\end{eqnarray}
Therefore, we conclude that $R = 0 = H$, i.e. Minkowski space, is a critical point. 

As anticipated, there is also a critical point at infinity, whose nature can be studied by using suitable a Poincar\'e's sphere transformation. This point at infinity represents a de Sitter repeller and thus provides a transient phase of Inflation at high-energy scales, i.e. for $\alpha R \gg 1$. The phase diagram for Starobinsky's model is displayed in figure \ref{Starmodel}.

\section{An exponential extension of the Starobinsky model}

Since Starobinsky's model is very successful, it is interesting to investigate small deviations from it in order to test its robustness. 
What we learned from the dynamical system analysis of the previous section is that $f(R)$ should be quadratic only for large values of $R$. Therefore, we propose the following model:
\begin{eqnarray}\label{expmodel}
	f(R) = R + \alpha R^2 - 2\Lambda e^{-\alpha R}\;,
\end{eqnarray}
i.e. an exponential correction of Starobinsky's model. 
For $\alpha R \gg 1$ our model reproduces the successful inflationary paradigm of Starobinsky's model and when $\alpha R \ll 1$, the above function $f(R)$ can be approximated as
\begin{eqnarray}\label{smallcurv}
	f(R) \sim R - 2\Lambda\;.
\end{eqnarray}

\begin{figure}[!tbp]
	\centering
	\subfloat[]{\includegraphics[width=0.3\textwidth]{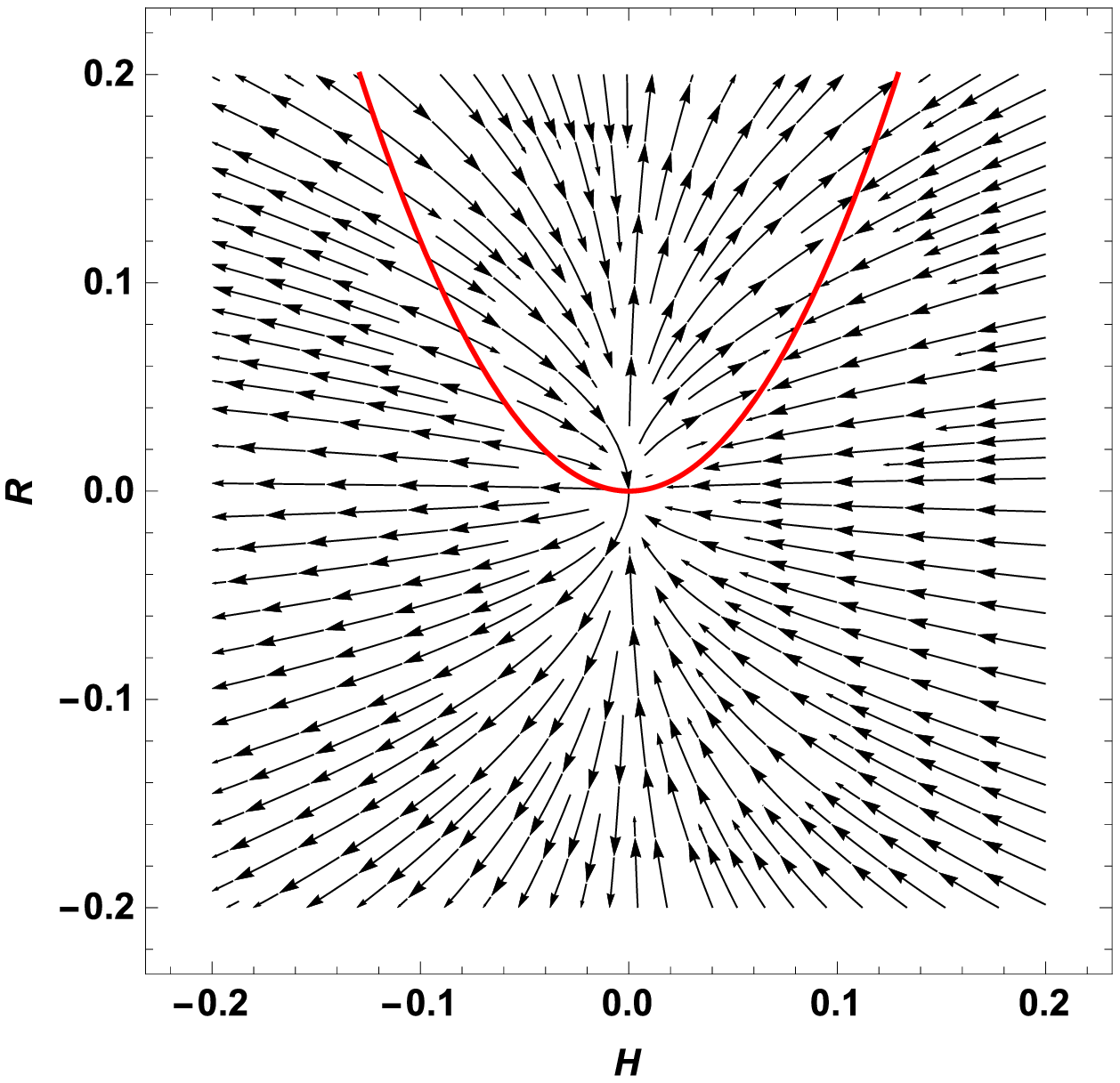}\label{R2}}
	\hfill
	\subfloat[]{\includegraphics[width=0.3\textwidth]{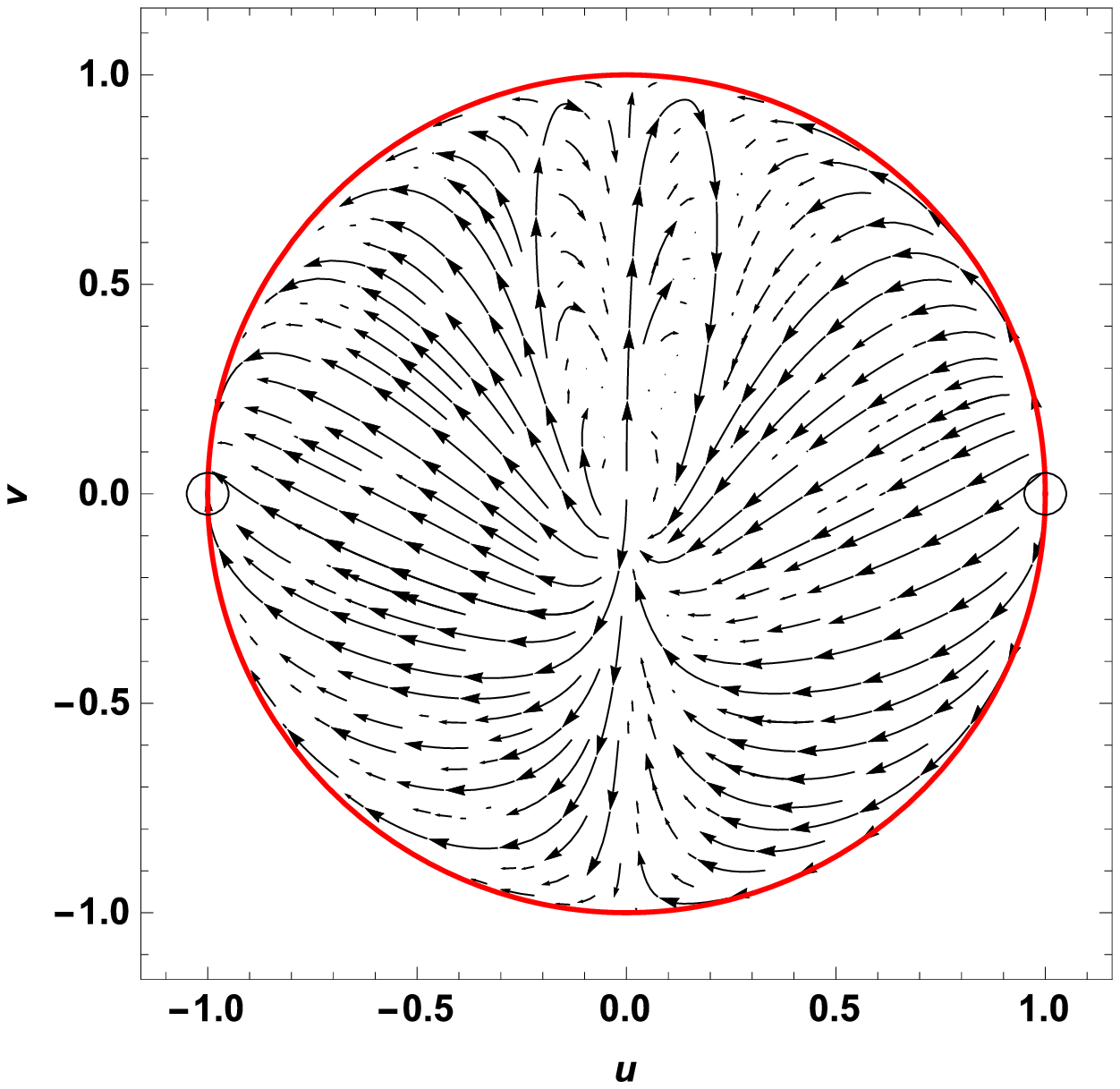}\label{Starmodel}}
	\hfill
	\subfloat[]{\includegraphics[width=0.3\textwidth]{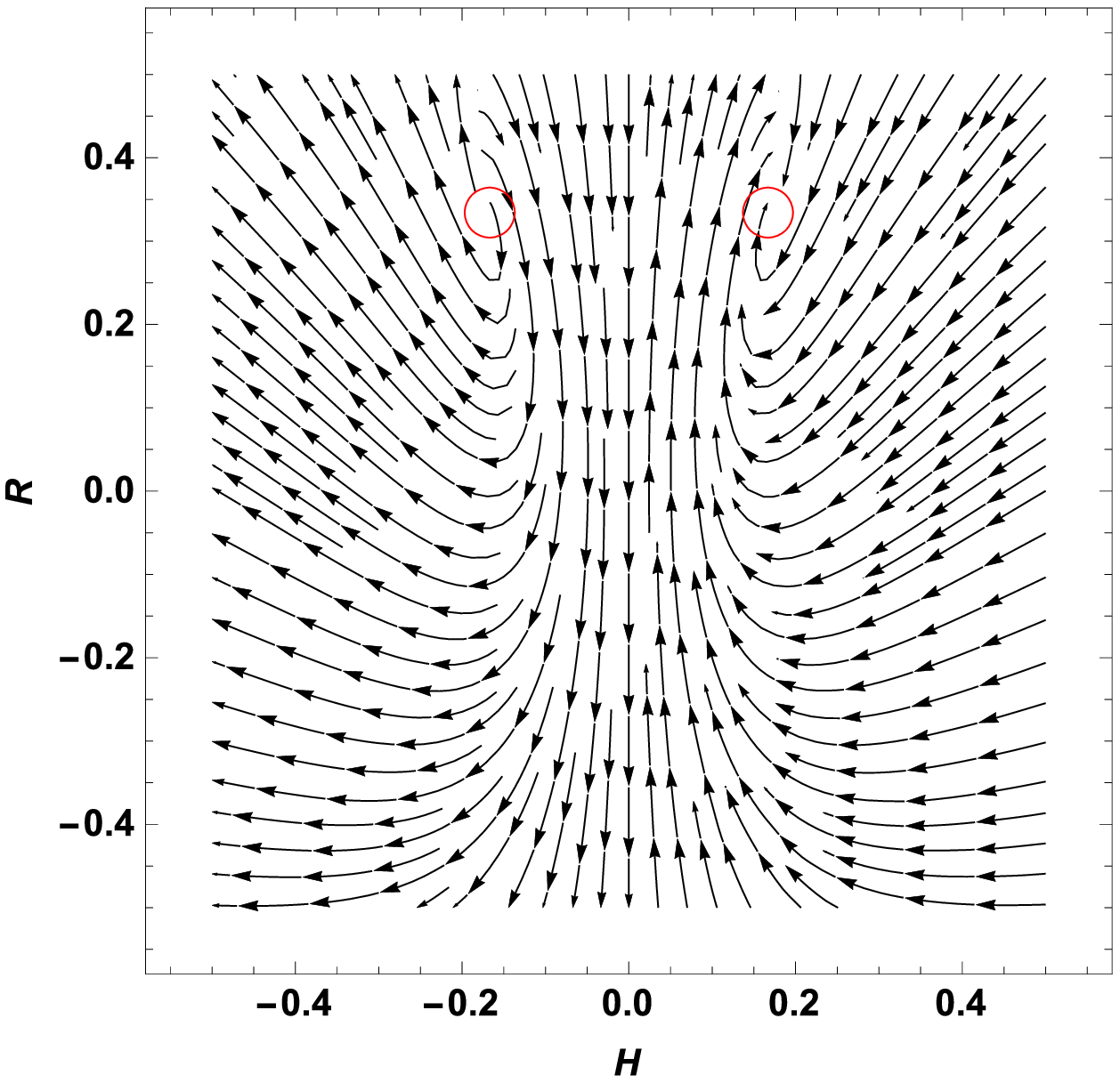}\label{Exp}}
	\caption{(a) Phase diagram for the $R^2$ model; (b) Equatorial plane of the Poincar\'e's sphere for the $f(R) = R + \alpha R^2$ model; (c) Phase space diagram for the $f(R) = R + \alpha R^2 - 2\Lambda e^{-\alpha R}$ model;}
\end{figure}

The dynamical system becomes:
\begin{eqnarray}\label{RdotHdotExpmodel}
	\begin{cases}
		\dot{R} = \frac{1}{12H\alpha}\frac{\alpha R^2 + 2\Lambda(\alpha R + 1)e^{-\alpha R} - 6H^2 - 12H^2\alpha R - 12H^2\alpha\Lambda e^{-\alpha R}}{1 - \alpha\Lambda e^{-\alpha R}}\;,\\
\dot{H} = \frac{R}{6} - 2H^2\;.
	\end{cases}
\end{eqnarray}
Looking for critical points, we get the same ones as in Starobinsky's model at infinity, representing inflation, and other new ones at the finite region given by the following transcendental equation:
\begin{eqnarray}\label{transceqexpmodel}
	R = 2\Lambda(\alpha R + 2)e^{-\alpha R}\;.
\end{eqnarray}
For $\alpha R \ll 1$, the above equation can be solved approximately as:
\begin{eqnarray}
	R \approx \frac{4\Lambda}{1 + 2\alpha\Lambda}\;. 
\end{eqnarray}
The phase diagram for this extended Starobinsky model is displayed in figure \ref{Exp}.

Let us inspect better such exponential extension of the Starobinsky model. A given $f(R)$ theory can be mapped into a scalar-tensor theory in the Jordan frame. Then, upon the conformal transformation $\hat{g}_{\mu\nu} = \kappa\phi g_{\mu\nu}$, we can write the action in the Einstein frame as follows:
\begin{eqnarray}\label{EinsteinFramef}
	S = \int d^4x\sqrt{-g}\left[\frac{1}{2\kappa^2}\hat{R} - \frac{1}{2}\hat{g}^{\mu\nu}\partial_\mu\chi\partial_\nu\chi - U(\chi)\right]\;,
\end{eqnarray}
where
\begin{eqnarray}\label{Uchirel}
	U \equiv \frac{V}{\kappa^2\phi^2}\;, \qquad \chi \equiv \sqrt{\frac{3}{2}}\frac{1}{\kappa}\ln\kappa\phi\;.
\end{eqnarray}

For the exponential model \eqref{expmodel}, the above defined quantities take the following form:
\begin{eqnarray}\label{phiRVRrel}
	\kappa\phi = 1 + 2\alpha R + 2\alpha\Lambda e^{-\alpha R}\;, \qquad 2\kappa^2V = \alpha R^2 + 2\Lambda(\alpha R + 1)e^{-\alpha R}\;. 
\end{eqnarray}

Finding an explicit form $V(\phi)$ for the potential in terms of elementary functions seems not possible. 
Formally, using the Lambert $W$ function (or product logarithm), we can write:
\begin{eqnarray}\label{Rphirel}
	2\alpha R = \kappa\phi - 1 + 2W\left[-\alpha\Lambda e^{(1-\kappa\phi)/2}\right]\;.
\end{eqnarray} 

The more intriguing feature of the potential of our exponential model, compared with the Starobinsky's model's one, is the presence of a minimum for a non-vanishing value of the scalar field near the origin. 
This suggests that the reheating phase of the Universe takes place earlier, if compared with the Starobinsky's model.
 
Let us now investigate the slow-roll parameters. They can be defined using the potential $U(\chi)$ in the definitions of eq.~(\ref{Slow}). The end of the inflationary period, i.e. when $\epsilon_U \approx 1$, takes place for smaller values of the field $\chi$ in the exponential extension of Starobinsky's model.

The number of e-foldings is defined as:
\begin{eqnarray}\label{efoldsnum}
	N \equiv \kappa^2\int_{\chi_f}^\chi\frac{U}{U_{,\chi}}d\chi\;,
\end{eqnarray} 
where $\chi_f$ is the scalar field at which inflation ends. Usually, that is determined by the condition $\epsilon_U(\chi_f) = 1$. The observables quantities are given by eq.(\ref{param}) and the the tensor-to-scalar power ratio $r=-8n_t$.

\begin{figure}[!tbp]
	\centering
	\subfloat[]{\includegraphics[width=0.3\textwidth]{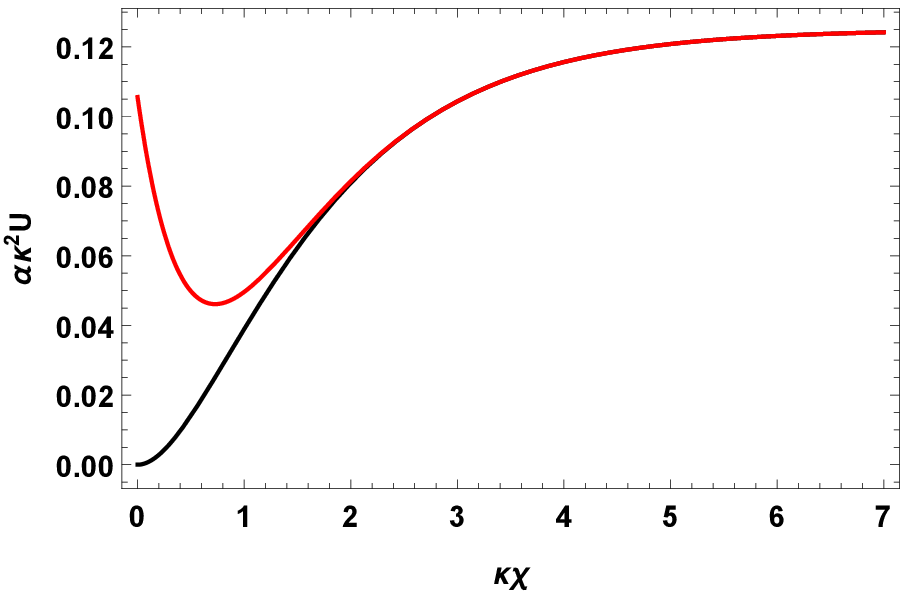}\label{Uchiplot}}
	\hfill
	\subfloat[]{\includegraphics[width=0.3\textwidth]{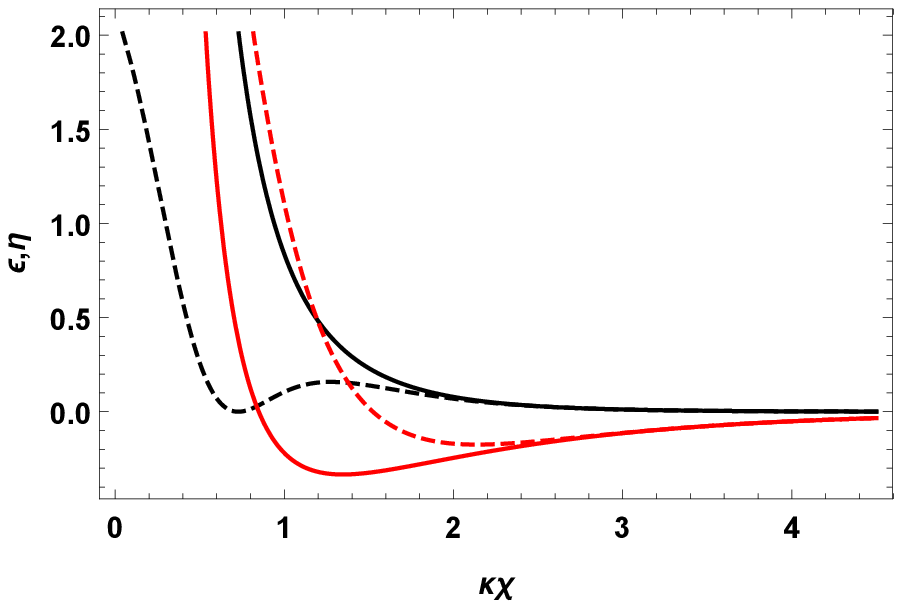}\label{slow}}
	\hfill
	\subfloat[]{\includegraphics[width=0.3\textwidth]{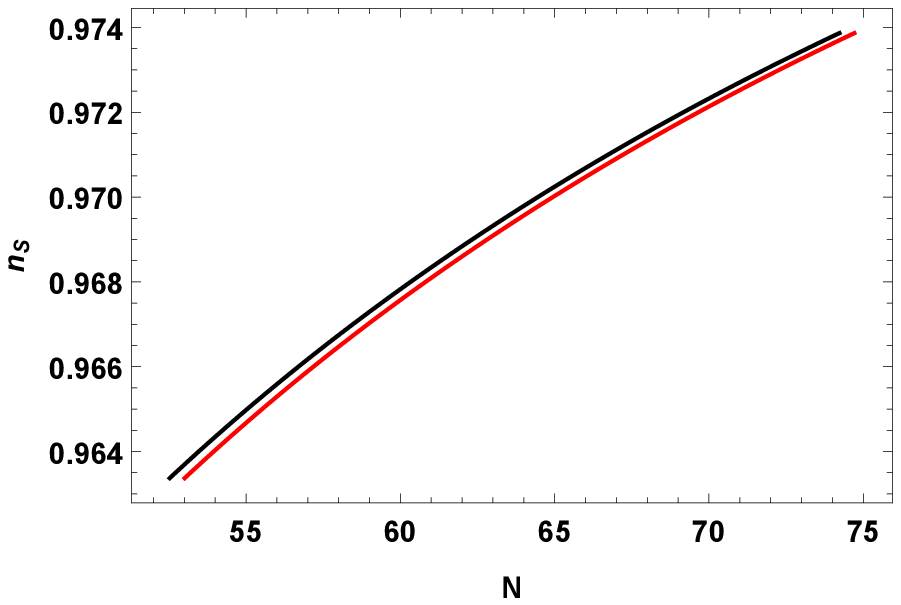}\label{epsilonNplot}}
	\caption{(a)Evolution of the effective potential $U(\chi)$, the black solid line represents Starobinsky's model (i.e. $\Lambda = 0$), whereas the red-line is its exponential extension for $\alpha\Lambda = 0.1$; (b)Evolution of the slow-roll parameters $\epsilon_U$ (black lines) and $\eta_U$ (red lines), the solid lines represent Starobinsky's model (i.e. $\Lambda = 0$), whereas the dashed lines are their exponential extensions for $\alpha\Lambda = 0.1$; (c)Evolution of the spectral index $n_s$ as function of the e-fold number $N$, the black solid line represents Starobinsky's model (i.e. $\Lambda = 0$), whereas the red-line is its exponential extension for $\alpha\Lambda = 0.1$;}
\end{figure}

The Planck collaboration put the following constraints on these quantities \cite{planck1, planck2}:
\begin{eqnarray}
	n_s = 0.968 \pm 0.006\;, \qquad r < 0.12\;,
\end{eqnarray}
the former being at 68\% CL and the latter at 95\% CL.
We can verify from figure \ref{epsilonNplot} that the predictions for the the extended Starobinsky model are very near the original model, what assures a good agreement with the observational test. However, it is important to investigate the production of gravitational waves in such extended model, in order to verify if
it can increase the possible values of the parameter $r$ with respect to the predictions of the original model.

\section{Discussion and conclusions}

Starobinsky's model of inflation is very successful and fits very well the observational data. One of the characteristic of such model is the very low production of gravitational waves, since the parameter $r$, measuring the ratio between the tensorial and scalar contribution to the anisotropy of the cosmic microwave background radiation is very small. A possible future detection of primordial gravitational waves may imply the falsification of Starobinsky's model. 

We have studied an extension of the Starobinsky model by adding an exponential term in the Lagrangian such that the correct behaviour of the original model is preserved, in particular the presence of critical points at the infinity region of the phase diagram. New features appear, essentially related to the appearance of new critical points at the finite region, representing a second de Sitter phase, which can be an attractive critical point. This creates the possibility to have a primordial inflationary phase, followed by "normal" phases in the evolution of the universe, and ending in a new inflationary phase (present stage of accelerated expansion?).

The observational predictions of the extended Starobinsky model seem to fit observations. But, a detailed study of gravitational wave production in this new model is necessary and constitute an important constraint for it.

\bigskip
\noindent
{\bf Acknowledgements:} We thank CNPq (Brazil), CAPES (Brazil) and FAPES (Brazil) for partial financial support. J.C.F. thanks the organising of the ICPPA conference (Moscow, october 2016) for the very nice ambiance during this meeting.


\begin{thebibliography}{99}
\bibitem{mukha} V. Mukhanov, {\bf Physical foundations of cosmology}, Cambridge university press, Cambridge(2005).
\bibitem{gliner} E.B. Gliner, Zhurnal Eksperimentalnoi i Teoreticheskoi Phiziki (in Russian), {\bf 49}, 542(1965).
\bibitem{martin} J. Martin, C. Ringeval, R. Trotta and V. Vennin, JCAP {\bf 1403}, 039(2014).
\bibitem{planck1} P. A. R. Ade et al. [Planck Collaboration], Astron. Astrophys. {\bf 571}, A22(2014).
\bibitem{planck2} P. A. R. Ade et al. [Planck Collaboration], arXiv:1502.02114 [astro.CO].
\bibitem{tsujikawa} S. Tsujikawa, {\bf Distinguishing between Inflationary Models from Cosmic Microwave Background}, Progress of Theoretical and Experimental Physics, Oxford University Press (2014).
\bibitem{star1} A. A. Starobinsky, JETP Lett. 30 (1979) 682 [Pisma Zh. Eksp. Teor. Fiz. 30 (1979) 719].
\bibitem{star2} A. A. Starobinsky, Phys. Lett. B 91 (1980) 99. doi:10.1016/0370-2693(80)90670-X
\bibitem{star3} A. A. Starobinsky, Zh. Eksp. Teor. Fiz. 34 (1981) 460.
\bibitem{luca} L. Amendola and Sh. Tsujikawa, {\bf Dark energy: theory and observations}, Cambridge university press, Cambridge(2010).
\bibitem{F} A. De Felice and Sh. Tsujikawa, Living Rev. Rel. {\bf 13}, 3 (2010).
\bibitem{ds} G. Sansone e R Conti. {\bf Equazioni differenziali non lineari}, edizioni cremonese, Cremona (1956).
\end{thebibliography}
\end{document}